\documentclass[twocolumn,showpacs,amssymb,amsmath,aps,pra]{revtex4-1}
\usepackage{graphics}
\usepackage{bm}
\usepackage{amsmath}
\usepackage{amssymb}

\begin{document}

\title{Generalized conditional entropy in bipartite quantum systems}
\author{N. Gigena and R. Rossignoli}
\affiliation{
Departamento de F\'{\i}sica-IFLP,
Universidad Nacional de La Plata, C.C. 67, La Plata (1900), Argentina}

\begin{abstract}
We analyze, for a general concave entropic form, the associated
conditional entropy of a
quantum system $A+B$, obtained as a result of a local measurement on one of the
systems ($B$). This quantity is a measure of the average mixedness of $A$ after
such measurement, and its minimum over all local measurements is shown to be
the associated entanglement of formation between $A$ and a purifying third
system $C$. In the case of the von Neumann entropy, this minimum determines
also the quantum discord. For classically correlated states and mixtures of a
pure state with the maximally mixed state, we show that the minimizing
measurement can be determined analytically and is {\it universal}, i.e., the
same for all concave forms. While these properties no longer hold for general
states, we also show that in the special case of the linear entropy, an
explicit expression for the associated conditional entropy can be obtained,
whose minimum among projective measurements in a general qudit-qubit state can
be determined analytically, in terms of the  largest eigenvalue of a simple
$3\times 3$ correlation matrix. Such minimum determines the maximum conditional
purity of $A$,  and the associated minimizing measurement is shown to be also
{\it universal} in the vicinity of maximal mixedness. Results for $X$ states,
including typical reduced states of spin pairs in $XY$ chains at weak and strong
transverse fields, are also provided and indicate that the measurements
minimizing the von Neumann and linear conditional entropies are typically
coincident in these states, being determined essentially by the main
correlation. They can differ, however, substantially from that minimizing the
geometric discord.
\end{abstract}

\maketitle

\section{Introduction}

There is presently a great interest in the investigation of quantum
correlations in mixed states of composite quantum systems. While for pure
states such correlations can be identified with entanglement, the situation in
mixed states is more complex, as separable (non entangled) mixed states,
defined as convex mixtures of product states \cite{RW.89} (i.e., states which
can be generated by local operations and classical communication),  can still
exhibit signatures of quantum-like correlations, manifested for instance in a
non-zero quantum discord \cite{OZ.01,HV.01,Zu.03}. Interest on this quantity
has been enhanced by the existence of mixed state based quantum algorithms
\cite{KL.98} able to achieve an exponential speed-up over the corresponding
classical algorithm with vanishing entanglement \cite{DFC.05} yet finite
discord \cite{DSC.08}. Various operational interpretations and implications of
states with non-zero discord have been recently provided
\cite{Mo.12,SKB.11,PGA.11,GTA.13}.

The quantum discord for a bipartite system $A+B$ can be written  \cite{OZ.01} as
the minimum difference between two distinct quantum extensions of the classical
Shannon based {\it conditional} entropy $S(A|B)$ \cite{Wh.78}, one involving a
local measurement $M_B$ on one of the systems ($B$), over which the
minimization is to be performed, and the other the direct quantum version of the
classically equivalent expression $S(A,B)-S(B)$ (which becomes negative in pure
entangled states). While other measures of quantum correlations with similar
properties (like reducing to an entanglement measure for pure states and
vanishing just for classically correlated states) have been introduced
\cite{Zu.03,Mo.12,SKB.11,GTA.13,HH.05,Lu.08,Mo.10,DVB.10,RCC.10,BB.12,GA.12,
LFO.12,POS.13,AD.12}, the quantum discord has the special feature, due to its
definition through a conditional entropy, of being directly related with the
entanglement of formation between the unmeasured system and a third system
which purifies the whole system \cite{KW.04,CC.11,MD.11,FC.11}. Accordingly,
the measurement minimizing the quantum discord  can differ substantially from
those minimizing other measures such as the geometric discord
\cite{DVB.10,Mo.12}, which can be much more easily determined. The complex
minimization involved in the quantum discord has in fact limited its evaluation
to simple systems or special states and measurements
\cite{DSC.08,Mo.12,GP.10,AD.10,FF.10,AR.10,CRC.10,GA.11,LL.11,CM.13}.

The aim of this work is first to extend the concept of measurement dependent
conditional entropy to a general entropic form (or uncertainty measure) $S_f$
depending on an arbitrary concave function $f$ \cite{Wh.78,CR.02}. The ensuing
quantity $S_f(A|B_{M_B})$ provides a measure of the average conditional
mixedness of $A$ after a measurement at $B$, and allows to define an associated
generalized ``information gain'' or uncertainty reduction
$I_f(A|B_{M_B})=S_f(A)-S_f(A|B_{M_B})$, which is non-negative for any concave
$f$ and reduces to the associated entanglement entropy $S_f(A)$ in the case of
pure states. Such extension  differs then from other treatments
\cite{AR.13,ML.13,AR.01} dealing with the generalization of the measurement
independent von Neumann conditional entropy $S(A,B)-S(B)$. The minimum of the
present $S_f(A|B_{M_B})$ among all local measurements coincides with the
associated entanglement of formation (convex roof extension of the $S_f$
entanglement entropy) between $A$ and a purifying third system $C$, as will be
shown.

Such general formulation allows, first, to recognize some universal features of
the measurement dependent conditional entropy which do not depend on the choice
of entropic function $f$ and rely just on concavity. It also opens the way to
use simple entropic forms like the linear entropy $S_2(\rho)=1-{\rm
Tr}\,\rho^2$, trivially related with the purity $P(\rho)={\rm Tr}\,\rho^2$ and
lower bound to the von Neumann entropy, which can be more easily evaluated (it
does not require the eigenvalues of $\rho$) and can therefore help to determine
and understand the minimizing measurement of the von Neumann conditional
entropy and hence the quantum discord. Moreover, we will show that this entropy
determines the behavior of {\it all} entropies in the vicinity of the maximally
mixed state. The purity, and hence $S_2(\rho)$, is also more easily accessible
from the experimental side,  since it can be determined efficiently without
requiring a full state tomography \cite{PP.02}.

We first derive in sec.\ 2 the fundamental properties of $S_f(A|B_{M_B})$,
including its minimum in general classically correlated states and mixtures of
a pure state with the maximally mixed state, where the minimizing measurement
is shown to be {\it universal}, i.e., the same for any entropic form. The
formalism is then applied in sec.\ 3  to derive a closed expression for the
conditional $S_2$ entropy and discuss its fundamental properties, including its
minimum over projective measurements for a general $A$+qubit system, which is
shown to be determined by the largest eigenvalue of a simple $3\times 3$
contracted correlation matrix. This permits to easily recognize the minimizing
measurement and understand its behavior. Applications to general parity
preserving two-qubit states ($X$ states),   including mixtures of aligned
states and weakly correlated states, relevant for the description of pair
states in interacting $XY$ spin chains at weak and strong transverse fields,
are presented in sec.\ 4. These examples indicate a similar behavior (and
coincidence of the minimizing measurement) of the $S_2$ and von Neumann
conditional entropies for these states, even well beyond the vicinity of
maximal mixedness. Conclusions are finally given in sec.\ 5.

\section{Formalism}
\subsection{Generalized conditional entropy after a local measurement}
We consider a bipartite quantum system $A+B$ in an initial state $\rho\equiv
\rho_{AB}$, with reduced states $\rho_A={\rm Tr}_B\,\rho$, $\rho_B={\rm
Tr}_A\,\rho$. We assume a general positive operator valued local measurement
\cite{NC.00} $M_B$ on system $B$ is performed, defined by a set of operators
$M_j=I_A\otimes M_j^B$, $j=1,\ldots,j_m$, such that the state after outcome $j$
is proportional to $M_j\rho M_j^\dagger$. The positive semidefinite operators
\begin{equation}
\Pi_j=M_j^\dagger M_j=I_A\otimes \Pi_j^B \,,\label{Pij}\end{equation}
should then satisfy $\sum_j \Pi_j =I\equiv I_A\otimes I_B$.

The reduced state of $A$ after outcome  $j$ depends just on $\Pi_j$ and is
given by
 \begin{equation}
\rho_{A/\Pi_j}=p_j^{-1}{\rm Tr}_B\,\rho \Pi_j\,,\;\;p_j={\rm Tr}\,\rho\,
 \Pi_j\,,\label{rhoa}\end{equation}
where $p_j>0$ is the probability of such outcome. In order to quantify the
average uncertainty or mixedness of the state of $A$ after such measurement, we
will consider here the generalized conditional entropy
\begin{eqnarray}
S_f(A|B_{\{\Pi_j\}})=\sum_j p_j S_f(\rho_{A/\Pi_j})\,,\label{Sfc}
\end{eqnarray}
where
\begin{equation} S_f(\rho)={\rm Tr}\,f(\rho)\,,\end{equation}
represents a generalized entropic form or uncertainty measure
\cite{Wh.78,CR.02} (see \ref{ApA}). Here $f:[0,1]\rightarrow\Re$ is a smooth
strictly concave function satisfying $f(0)=f(1)=0$. For
$f(\rho)=-\rho\,\log_a\,\rho$ (we use here $a=2$ or $e$), $S_f(\rho)$ becomes
the von Neumann entropy $S(\rho)=-{\rm Tr}\,\rho\log_a\rho$, and Eq.\
(\ref{Sfc}) the measurement dependent von Neumann conditional entropy,
introduced in \cite{OZ.01} for the definition of the quantum discord.

The concavity of these forms, i.e.,
\begin{equation}
S_f(\sum_\alpha q_\alpha \rho_\alpha)
 \geq \sum_\alpha q_\alpha S_f(\rho_\alpha)\,,\label{conc}
 \end{equation}
if $\{q_\alpha\}$ is a probability distribution ($q_\alpha\geq 0$, $\sum_\alpha
q_\alpha=1$) and all $\rho_\alpha\,$'s are quantum states, directly follows
from the concavity of $f$, and implies fundamental properties of the
generalized conditional entropy (\ref{Sfc}). First, since $\rho_A=\sum_j
p_j\,\rho_{A/\Pi_j}$, Eq.\ (\ref{conc}) implies $S_f(A)\equiv S_f(\rho_A)\geq
\sum_j p_j S_f(\rho_{A/\Pi_j})$, i.e.,
\begin{equation}
S_f(A)\geq S_f(A|B_{\{\Pi_j\}})\,,\label{SfA}
\end{equation}
indicating that the average conditional mixedness of $A$ after a measurement at
$B$, will not exceed the original mixedness, for {\it any} choice of $S_f$.
Moreover, if $f$ is strictly concave, equality in (\ref{conc}) holds iff all
$\rho_\alpha$'s with $q_\alpha>0$ are identical. Hence, equality in (\ref{SfA})
for {\it all} $M_B$ holds just if $\rho=\rho_A\otimes \rho_B$, since only in
this case $\rho_{A/\Pi_j}=\rho_A$ $\forall$ $\Pi_j$.
The quantity
\begin{equation}
I_f(A|B_{\{\Pi_j\}})=S_f(A)-S_f(A|B_{\{\Pi_j\}})\,,\label{Sfm}\end{equation}
{\it is then non-negative for any} $S_f$, vanishing for all $M_B$ just for
product states. It represents the average reduction in the quantum uncertainty
of $A$ (or generalized information gain about $A$) as measured by $S_f$, after
a measurement at $B$.

Eq.\ (\ref{conc}) also implies concavity of the conditional entropy: If
$\rho=\sum_\alpha q_\alpha \rho^\alpha$, then $\rho_{A/\Pi_j}=\sum_\alpha
p_j^{-1} q_\alpha p_j^\alpha\,\rho^\alpha_{A/\Pi_j}$, with $p_j^\alpha={\rm
Tr}\rho^\alpha\Pi_j$, $p_j=\sum_\alpha q_\alpha p_j^\alpha$. Hence,
$S_f(\rho_{A/\Pi_j})\geq \sum_\alpha p_j^{-1}q_\alpha p_j^\alpha
S_f(\rho^\alpha_{A/\Pi_j})$, entailing
\begin{equation}
S_f(A|B_{\{\Pi_j\}})\geq \sum_\alpha q_\alpha
S_f(A^\alpha|B^\alpha_{\{\Pi_j\}})\,,
 \label{condcc}\end{equation}
where $S_f(A^\alpha|B^\alpha_{\{\Pi_j\}})=\sum_{j}p_j^\alpha
S_f(\rho^\alpha_{A/\Pi_j})$: Average uncertainty about $A$ after state mixing
cannot be smaller than the average of the original average uncertainties. In
addition, if
\begin{equation}\Pi_j=\sum_{k}r_j^k
 \tilde{\Pi}_k\,,\;\;\;r_j^k\geq 0\,,\label{SP}\end{equation}
where $\tilde{\Pi}_k=I_A\otimes\tilde{\Pi}_k^B$, with $\sum_k \tilde{\Pi}_k=I$,
are positive operators representing a more detailed local measurement (i.e.,
all $\tilde{\Pi}_k^B$ of rank 1) and $\sum_j r_j^k=1$, then
$\rho_{A/\Pi_j}=\sum_k p_j^{-1} r_j^k q_k\rho_{A/\tilde{\Pi}_k}$, with
$q_k={\rm Tr}\,\rho\tilde{\Pi}_k$,  $\sum_k r_j^k q_k=p_j$. Hence,
$S_f(\rho_{A/\Pi_j})\geq \sum_k p_j^{-1}r_j^k q_k S_f(\rho_{A/\tilde{\Pi}_k})$
and
\begin{equation}S_f(A|B_{\{\Pi_j\}})\geq \sum_{k} q_k S_f(\rho_{A/\tilde{\Pi}_k})
 =S_f(A|B_{\{\tilde{\Pi}_k\}})\,,\label{Sfjk}\end{equation}
i.e., the generalized conditional entropy will not increase (and will in
general decrease) if a more detailed local measurement is performed. In fact,
$S_f(A)$ can be considered as the conditional entropy $S_f(A|B_{I})$ of $A$
after a trivial measurement of the identity $I_B$ in $B$, so that Eq.\
(\ref{SfA}) is a particular case of (\ref{Sfjk}).

Minimum uncertainty about the state of $A$ will then be obtained for
measurements based on rank one operators
\begin{equation}\Pi_j^B=r_j
 |j_B\rangle\langle j_B|\,,\;\;r_j>0\,,\label{M1}\end{equation}
where $|j_B\rangle$ are normalized states such that $\sum_j \Pi_j^B=I_B$.
Standard complete projective measurements (von Neumann measurements) correspond
to $r_j=1$ and $\{|j_B\rangle\}$ an orthonormal basis
($\Pi_j\Pi_{j'}=\delta_{jj'}\Pi_j$). In particular, for pure states
$\rho^2=\rho$, i.e.,
\begin{equation}
\rho=|\Psi\rangle\langle\Psi|,\;\;|\Psi\rangle=\sum_k\sqrt{q_k}
\,|\tilde{k}_A\tilde{k}_B\rangle\,,
 \label{SB}\end{equation}
where the last expression denotes the Schmidt decomposition \cite{NC.00}
($\{|k_A\rangle\}$, $\{|k_B\rangle\}$ orthonormal sets), $\rho_{A/\Pi_j}$ is
{\it pure} $\forall$ $j$ with $p_j>0$,  for {\it any} local measurement based
on the operators (\ref{M1}):
\begin{equation}
\rho_{A/\Pi_j}=|j_A\rangle\langle j_A|,\;\;
|j_A\rangle=
(r_j/p_j)^{1/2}\sum_{k}\sqrt{q_k}\langle j_B|\tilde{k}_B\rangle
|\tilde{k}_A\rangle\,,
\label{ja}
\end{equation}
where $p_j=r_j\sum_k q_k|\langle j_B|\tilde{k}_B\rangle|^2$. Hence, in the pure
case $S_f(A|B_{\{\Pi_j\}})=0$, and Eq.\ (\ref{Sfm}) becomes the
generalized entanglement entropy \cite{RCC.10}:
\begin{equation}I_f(A|B_{\{\Pi_j\}})=S_f(A)=S_f(B)=\sum_k f(q_k)
\label{Sfp}\,.\end{equation}

\subsection{Minimum  conditional entropy and generalized entanglement
of formation}
Let us now consider the minimum of Eq.\ (\ref{Sfc}) among all local
measurements $M_B$ for a general state $\rho$,
\begin{equation}
S_f(A|B)\equiv\mathop{\rm Min}_{\{\Pi_j\}}S_f(A|B_{\{\Pi_j\}})\,.
 \label{Sm}\end{equation}
From Eq.\ (\ref{Sfjk}) it follows that just rank one operators of the form
(\ref{M1}) need to be considered in the minimization. Eq.\ (\ref{Sm}) leads to
the maximum generalized information gain (i.e., maximum uncertainty reduction)
\begin{equation}
I_f(A|B)=\mathop{\rm Max}_{\{\Pi_j\}}I_f(A|B_{\{\Pi_j\}})=S_f(A)-S_f(A|B)\,.
\label{IM}
\end{equation}

If the system $A+B$ is purified \cite{NC.00} by a adding a third system $C$,
Eq.\ (\ref{Sm}) has the important meaning of being the associated {\it
entanglement of formation}  $E_f(A,C)$ \cite{RCC.10} between $A$ and $C$ in the
reduced state $\rho_{AC}$ \cite{KW.04}:
\begin{equation}
S_f(A|B)=E_f(A,C)=\mathop{\rm Min}_{\sum_j p_j
\rho^j_{AC}=\rho_{AC}}\sum_j p_j S_f(
\rho^j_{A})\,,\label{idf}
\end{equation}
where the minimization is over all representations of $\rho_{AC}$ as convex
combination ($p_j> 0$) of pure states $\rho^j_{AC}=|j_{AC}\rangle\langle
j_{AC}|$, and $S_f(\rho_A^j)=S_f(\rho_C^j)$ is the $S_f$ entanglement entropy
between $A$ and $C$ in $|j_{AC}\rangle$ ($\rho_{A}^j={\rm Tr}_C\,\rho^j_{AC}$).
Eq.\ (\ref{idf}) is the convex roof extension \cite{Ca.03} of the pure state
entanglement entropy (\ref{Sfp})  and is an entanglement monotone \cite{Vi.00}.
The identity (\ref{idf}) was derived for the von Neumann entropy (see
\cite{KW.04} and \cite{CC.11,MD.11,FC.11}), where $E_f(A,C)$ becomes the
standard entanglement of formation $E(A,C)$ \cite{EFO.96}, but the arguments
remain valid in the present general case (see  \ref{ApB}).

Eq.\ (\ref{idf}) entails that the Eq.\ (\ref{IM}) can be
also expressed as
\begin{equation}
I_f(A|B)=E_f(A,BC)-E_f(A,C)\,,
\end{equation}
where $E_f(A,BC)=S_f(\rho_A)=S_f(\rho_{BC})$ is the entanglement entropy
between $A$ and $BC$ in the purified state.

The quantum discord \cite{OZ.01,HV.01,Zu.03,Mo.12} $D(A|B)$, as obtained by a
measurement in $B$, is directly related to the present  {\it von
Neumann} conditional entropy $S(A|B_{\{\Pi_j\}})$ through
\begin{equation} D(A|B)=\mathop{\rm Min}_{\{\Pi_j\}}
S(A|B_{\{\Pi_j\}})-[S(A,B)-S(B)]\,,
\end{equation}
where the last bracket is the standard (measurement independent) quantum
extension of the von Neumann conditional entropy (which can be negative in
entangled states). It can be also expressed as the difference between the
standard mutual information $S(A)+S(B)-S(A,B)$ and the maximum von Neumann
information gain $I(A|B)=S(A)-{\rm Min}_{\{\Pi_j\}}S(A|B_{\{\Pi_j\}})$. A
generalization of the quantum discord based on the Renyi entropy of order $2$
was considered in \cite{AD.12} for gaussian states, whereas extensions based on
the Tsallis entropy \cite{Ts.09}were discussed in \cite{TQD.12}.

\subsection{Classically correlated states}
There are important classes of mixed states where the local measurement
minimizing $S_f(A|B_{\{\Pi_j\}})$ is {\it universal}, i.e., the same for {\it
all} entropies $S_f$, and can be generally determined. One is that of
classically correlated states with respect to $B$ \cite{OZ.01,HV.01,Zu.03},
\begin{equation}
\rho=\sum_k q_k\rho_{A/k}\otimes \tilde{\Pi}_k^B\,,
\label{rk}
\end{equation}
where $q_k\geq 0$ and $\{\tilde{\Pi}_k^B=|\tilde{k}_B\rangle
\langle\tilde{k}_B|\}$ is a complete set of orthogonal rank one local
projectors, such that after a local measurement in this basis,
$\rho_{A/\tilde{\Pi}_k}=\rho_{A/k}$ (and $\sum_k\tilde{\Pi}_k\rho
\tilde{\Pi}_k=\rho$ if $\tilde{\Pi}_k=I_A\otimes\tilde{\Pi}_k^B$, implying that
the states (\ref{rk}) remain unchanged after an unread local measurement in
this basis). It is easy to prove that the lowest conditional entropy (\ref{Sm})
is obtained for such measurement, for {\it any} $S_f$:
\begin{equation}
S_f(A|B)=S_f(A|B_{\{\tilde{\Pi}_k\}})=\sum_k q_k S_f(\rho_{A/k})\,.
\end{equation}
Proof: For any $M_B$ based on the
operators (\ref{M1}), we have
\begin{equation}
\rho_{A/\Pi_j}=\sum_k r_j p_j^{-1}q_k |\langle
j_B|\tilde{k}_B\rangle|^2\,\rho_{A/k}\,,
\end{equation}
with $p_j=r_j\sum_k q_k|\langle j_B|\tilde{k}_B\rangle|^2$. Concavity
plus completeness ($\sum_j r_j |\langle j_B|\tilde{k}_B\rangle|^2=1$) imply
\begin{eqnarray}
S_f(A|B_{\{\Pi_j\}})&\geq& \sum_{k,j}r_jq_k
|\langle j_B|\tilde{k}_B\rangle|^2S_f(\rho_{A/k})\nonumber\\
&=&\sum_k q_k S_f(\rho_{A/k})\,, \label{ineqp}
\end{eqnarray}
with the inequality saturated for a measurement in the pointer basis
$\{|\tilde{k}_B\rangle\}$, formed by the eigenstates of $\rho_B=\sum_k
q_k\tilde{\Pi}_k^B$. The maximum $I_f$ is then
\begin{equation}
I_f(A|B)=S_f(\sum_k q_k \rho_{A/k})-\sum_k q_k S_f(\rho_{A/k})\,.
\end{equation}

\subsection{Pure state plus maximally mixed state}
A second case is that of the mixture of a general pure state (\ref{SB}) with
the maximally mixed state $I/d$,
\begin{equation}
\rho=w|\Psi\rangle\langle\Psi|+(1-w)I_d/d\,,
\;\;\;\;\;|\Psi\rangle=\sum_k\sqrt{q_k}
\,|\tilde{k}_A\tilde{k}_B\rangle\,,\label{rw}
\end{equation}
where $w\in[0,1]$ and $d=d_A d_B$ is the Hilbert-space dimension of $A+B$. The
minimum for {\it any} $S_f$ is provided again by a measurement in the basis
$\{|\tilde{k}_B\rangle\}$ of eigenstates of $\rho_B$:
\begin{eqnarray}
 S_f(A|B)&=&S(A|B_{\{\tilde{\Pi}_k\}})
=\sum_k q_k^w
 S_f(\rho_{A/\tilde{\Pi}_k})\nonumber\\&=&
\sum_k q_k^w[{\textstyle f(\frac{w q_k+(1-w)/d}{q_k^w})+
 (d_A-1)f(\frac{1-w}{dq_k^w})]}\,, \nonumber\\&&\label{sfex}
\end{eqnarray}
where $q_k^w=wq_k+\frac{1-w}{d_B}$ is the probability of outcome $k$ at $B$ and
$\rho_{A/\tilde{\Pi}_k}=[w q_k|\tilde{k}_A\rangle\langle
\tilde{k}_A|+(1-w)I_A/d]/q_k^w$ the state of $A$ after such outcome.

Proof: For any measurement based on the operators (\ref{M1}) we obtain, using
(\ref{SB})--(\ref{ja}),
\begin{eqnarray}
\rho_{A/\Pi_j}&=&\frac{w p_j|j_A\rangle\langle
 j_A|+r_j(1-w)I_A/d}{p_j^w}\nonumber\\
&=&\sum_{k} \frac{r_jq_k^w}{p_j^w}
|\langle j_B|\tilde{k}_B\rangle^2|U_k^j\rho_{A/\tilde{\Pi}_k}{U_k^j}^\dagger\,,
\end{eqnarray}
where $p_j=r_j\sum_{k}q_k|\langle j_B|\tilde{k}_B\rangle|^2$ and
$p_j^w=wp_j+r_j\frac{1-w}{d_B}$  are respectively the probabilities of outcome
$j$ in $|\Psi\rangle$ and $\rho$, and $U_k^j$ are unitaries satisfying
$U_k^j|\tilde{k}_A\rangle=|j_A\rangle$. Hence, concavity, invariance of $S_f$
under unitary transformations and completeness imply again
\begin{eqnarray}
S_f(A|B_{\{\Pi_j\}})&\geq&
\sum_k q_k^w S_f(\rho_{A/\tilde{\Pi}_k})=S_f(A|B_{\{\tilde{\Pi}_k\}})\,.\label{Skk}
\end{eqnarray}
Equality in (\ref{Skk}) for any $M_B$ of the form (\ref{M1}) holds
for i) $w=0$ ($\rho$ maximally mixed), ii) $w=1$ ($\rho$ pure) and iii)
$|\Psi\rangle$ {\it maximally entangled} ($q_k=1/d_B$ $\forall$ $k$, assuming
$d_A\geq d_B$), where $p_j=r_j/d_B$ $\forall$ $j$ and  all
$\rho_{A/\Pi_j}=w|j_A\rangle\langle j_A|+
\frac{1-w}{d_A}I_A$ have the same spectrum.

It can be easily checked that Eq.\ (\ref{sfex}) is a {\it concave} function of
both $w$ and the probability distribution $\bm{q}=\{q_k\}$. Since $S_f(A|B)$
reaches its maximum $S_f(I_A/d_A)=d_Af(1/d_A)$ for $w=0$,  concavity entails
that Eq.\ (\ref{sfex}) is a {\it decreasing} function of $w$ for $w\in[0,1]$
$\forall$ $S_f$: Decreasing mixedness decreases the uncertainty about $A$.
Concavity also leads to the immediate lower bound $S_f(A|B)\geq
(1-w)d_Af(1/d_A)$.

Besides, for states $|\Psi\rangle$, $|\Psi'\rangle$ characterized by
distributions $\bm{q}$ and $\bm{q}'$ in the Schmidt decomposition, we have
$S_f(\bm{q})\geq S_f(\bm{q}')$ $\forall$ $S_f$ iff $\bm{q}\prec \bm{q}'$ (i.e.,
$\bm{q}$ majorized by $\bm{q}'$, see \ref{ApA}). Such condition ensures then
that $|\Psi\rangle$ {\it is more entangled than $|\Psi'\rangle$ for any $S_f$},
and is the same condition which warrants that $|\Psi'\rangle$ can be obtained
from $|\Psi\rangle$ by LOCC \cite{NC.00,Ni.99}.  In such a case, concavity of
$S_f(A|B)$ with respect to $\bm{q}$ entails that at fixed $w\in(0,1)$,
$S_f(A|B)_{|\Psi\rangle}\geq S_f(A|B)_{|\Psi'\rangle}$ for {\it any} $S_f$,
i.e., greater entanglement for any $S_f$ entails a larger conditional entropy
$S_f(A|B)$ $\forall$ $S_f$ in the mixture (\ref{rw}), in contrast with the pure
case $w=1$ (where $S_f(A|B)=0$ for any pure state $|\Psi\rangle$).

\section{The quadratic case: Conditional purity after  local measurement}
\subsection{General properties}
We now consider in detail the simplest choice of concave $f$, i.e., a quadratic
function  $f(\rho)=\alpha(\rho-\rho^2)$, $\alpha>0$. For $\alpha=1$ this leads
to $S_f(\rho)=S_2(\rho)$, with
\begin{equation}
S_2(\rho)=1-{\rm Tr}\,\rho^2\,,\label{S2}
\end{equation}
the so called {\it linear entropy}, since it corresponds to the linear
approximation $-\ln \rho\approx I-\rho$ in $S(\rho)$ ($\ln p=p-1+O(p-1)^2$ for
$p\rightarrow 1$). It is the $q=2$ case of the Tsallis entropy $S_q(\rho)$
\cite{Ts.09} (see \ref{ApA})  and provides a lower bound to the von Neumann
entropy for $a=e$ (and hence $a<e$), since $p(1-p)\leq -p\ln p$ $\forall$
$p\in[0,1]$.

Eq.\ (\ref{S2}) is trivially related with the {\it purity} $P(\rho)={\rm
Tr}\,\rho^2$, which satisfies $P(\rho)\leq 1$, with $P(\rho)=1$ iff $\rho$ is a
pure state ($\rho^2=\rho$). It is also directly related to the squared
Hilbert-Schmidt distance to the maximally mixed state $I/d$:
\begin{equation}
||\rho-I/d||^2={\rm Tr}\,\rho^2-1/d=S_2(I/d)-S_2(\rho)\,,
\label{S2I}
\end{equation}
where $||O||^2={\rm Tr}\,O^\dagger O$ and $S_2(I/d)=1-1/d$.

Similarly, the associated conditional entropy
\begin{equation} S_2(A|B_{\{\Pi_j\}})=1-\sum_j p_j\,
 {\rm Tr}\, \rho^2_{A/\Pi_j}\,,\end{equation}
is trivially related with the average conditional purity
$P(A|B_{\{\Pi_j\}})=\sum_j p_j {\rm Tr}\,\rho^2_{A/\Pi_j}$, and determines the
average squared distance to the maximally mixed state of $A$:
\begin{equation}
\sum_j p_j||\rho_{A/\Pi_j}-I_A/d_A||^2=S_2(I_A/d_A)-S_2(A|B_{\{\Pi_j\}})
\,.\label{S2IA}
\end{equation}
The ensuing $I_2(A|B)$  represents the average increase of
the purity of $A$ due to the local measurement at $B$, and can be also
interpreted as the average squared distance between the original and the
post-measurement state of $A$:
\begin{eqnarray}
I_2(A|B_{\{\Pi_j\}})&=&
S_2(A)-S_2(A|B_{\{\Pi_j\}})\nonumber\\&=&\sum_j p_j {\rm Tr}\rho_{A/\Pi_j}^2
-{\rm Tr}\rho_A^2\,\label{I21}\\
&=&\sum_j p_j\,||\rho_A-\rho_{A/\Pi_j}||^2\,,\label{I22}
\end{eqnarray}
where  we used Eq.\ (\ref{rhoa}).  We may also define, through $I_2$ and $S_2$,
the purity gain ratio
\begin{equation}
R_2(A|B_{\{\Pi_j\}})=1+\frac{I_2(A|B_{\{\Pi_j\}})}{1-S_2(A)}=\frac{\sum_j p_j
{\rm Tr}\rho_{A/\Pi_j}^2} {{\rm Tr}\rho_A^2}\,,
\end{equation}
which satisfies $1\leq R_2(A|B_{\{\Pi_j\}})\leq d_A$. Such ratio remains
unaltered if an ancilla $C$ at $A$ is added
($\rho_{AB}\rightarrow\rho_C\otimes\rho_{AB}$).

If $\rho$ is sufficiently close to the maximally mixed state $I/d$, Eq.\
(\ref{S2I}) entails that {\it all} entropies $S_f(\rho)$ (with $f''(p)<0$
$\forall$ $p$) become in this limit {\it linear functions} of $S_2(\rho)$: A second
order expansion of $S_f(\rho)$ around $\rho=I/d$ leads to
\begin{eqnarray}
 S_f(\rho)-S_f(I/d)&\approx&
 {\textstyle\frac{1}{2}f''(\frac{1}{d})||\rho-I/d||^2}
 \nonumber\\ &=&{\textstyle \frac{1}{2}|f''(\frac{1}{d})|
[S_2(\rho)-S_2(I/d)]}\,.
\end{eqnarray}
Hence, in the vicinity of maximal mixedness, {\it all} entropies $S_f(\rho)$
(with $f''(1/d)<0$), including of course the von Neumann entropy $S(\rho)$, are
determined by $S_2(\rho)$. In this limit $\rho_{A/\Pi_j}$ is also close to
$I_A/d_A$ $\forall$ $\Pi_j$ and hence,
\begin{eqnarray}
S_f(A|B_{\{\Pi_j\}})&\approx& S_f(I_A/d_A)+\nonumber\\&&
{\textstyle\frac{1}{2}|f''(\frac{1}{d_A})|}[S_2(A|B_{\{\Pi_j\}})-S_2(I_A/d_A)]\,,
\end{eqnarray}
indicating that  {\it all} conditional entropies $S_f(A|B_{\{\Pi_j\}})$ (with
$f''(1/d_A)<0$) also become functions of the $S_2$ conditional entropy. The
measurement minimizing $S_2(A|B_{\{\Pi_j\}})$ becomes then {\it universal} in
this limit, i.e., it will also minimize all other $S_f(A|B_{\{\Pi_j\}})$.

We note here that the {\it geometric discord} \cite{DVB.10,Mo.12} is defined as
the minimum squared Hilbert-Schmidt distance from $\rho$ to a classically
correlated state $\rho_c$ of the form (\ref{rk}), and is equivalent to the
minimum increase of the $S_2$ entropy of the global state due to an unread
projective measurement at $B$ \cite{RCC.10}:
\begin{equation}D_2(A|B)=\mathop{\rm Min}_{\rho_c}||\rho-\rho_c||^2=
{\rm Min}_{\{\Pi_j\}}\,S_2(\sum_j
 \Pi_j\rho\Pi_j)-S_2(\rho)\,,\label{D2}\end{equation}
where again $\Pi_{j}=I_A\otimes\Pi_{j}^B$.  In contrast with $S_2(A|B)$, the
geometric discord looks for the closest average global post-measurement state
$\sum_j \Pi_j\rho\Pi_j$. This will lead to significant differences in the
minimizing measurement for certain states, as discussed in sec.\ 4.

\subsection{Explicit expressions}
The obvious advantage of $S_2(\rho)$ over other entropies is that its
evaluation does not require the knowledge of the eigenvalues of $\rho$.
Convenient expressions in a system with Hilbert space dimension $d$ can be
obtained just by considering a complete  orthogonal set of hermitian operators
$(I,\bm{\sigma})$, with $\bm{\sigma}=(\sigma_1,\ldots,\sigma_{d^2-1})$
satisfying
\begin{equation}
{\rm Tr}\, \sigma_i=0,\;\; {\rm
 Tr}\,\sigma_i\sigma_j=d\delta_{ij}\,.\end{equation}
For a single qubit $\bm{\sigma}$ are the Pauli operators. A general state can
then be written as
\begin{equation}
\rho=(I+\bm{r}\cdot\bm{\sigma})/d\,,\;\;
\bm{r}={\rm Tr}\,\rho\bm{\sigma}=\langle\bm{\sigma}\rangle\,,\label{rhx}
\end{equation}
and the quadratic entropy (\ref{S2}) becomes
\begin{equation}
S_2(\rho)=1-(1+|\bm{r}|^2)/d.\label{S2r}
\end{equation}
For a pure state $\rho^2=\rho$, $|\bm{r}|^2=d-1$ and $S_2(\rho)=0$.

In the case of a bipartite system $A+B$, we may rewrite Eq.\ (\ref{rhx}) as
\begin{eqnarray}
\rho&=&[I+\bm{r}_A\cdot\bm{\sigma}_A\otimes I_B+
I_A\otimes\bm{r}_B\cdot\bm{\sigma}_B+\bm{\sigma}_A^tJ\otimes \bm{\sigma}_B]/d
\nonumber\\&&
\label{rhod}
\end{eqnarray}
where $\bm{r}_{A}=\langle \bm{\sigma}_{A}\rangle$,
$\bm{r}_B=\langle\bm{\sigma}_B\rangle$ and $J=\langle\bm{\sigma}_A\otimes
\bm{\sigma}_B^t\rangle$ is a $(d_A^2-1)\times (d_B^2-1)$ matrix of elements
$J_{ij}=\langle\sigma_{Ai}\otimes \sigma_{B j}\rangle$. The reduced states are
$\rho_\alpha=(I_\alpha+\bm{r}_\alpha\cdot\bm{\sigma}_\alpha)/d_\alpha$,
$\alpha=A,B$.

A measurement $M_B$ based on the operators (\ref{M1}) can be characterized by
the vectors
\begin{equation}\bm{k}_j={\rm Tr}_B (\bm{\sigma}_B\,|j_B\rangle\langle j_B|)\,,
\end{equation}
such that $\Pi_j^B=r_j(I_B+\bm{k}_j\cdot\bm{\sigma}_B)/d_B$. These vectors
satisfy $|\bm{k}_j|^2=d_B-1$ and
\begin{equation}
\sum_j r_j\bm{k}_j=\bm{0}\,,\label{0}
\end{equation}
since $\sum_j \Pi_j^B=I_B$.
The probability of outcome $j$ and the ensuing
state $\rho_{A/\Pi_j}$ are then
\begin{equation}
p_j=\frac{r_j}{d_B}(1+\bm{r}_B\cdot\bm{k}_j)\,,\;\;
\rho_{A/\Pi_j}=
\frac{1}{d_A}\left[I_A+\frac{(\bm{r}_A+J\bm{k}_j)
\cdot\bm{\sigma}_A}{1+\bm{r}_B\cdot\bm{k}_j}\right]\,,\;\;
\label{roa2}
\end{equation}
which involve just the components of $\bm{r}_B$ and $J$ along $\bm{k}_j$.
Eqs.\ (\ref{S2r})--(\ref{roa2}) lead then to
\begin{eqnarray}S_2(A|B_{\{\Pi_j\}})&=&
1-\frac{1}{d_A}\left[1+\sum_j p_j\frac{|\bm{r}_A+
J\bm{k}_j|^2}{(1+\bm{r}_B\cdot\bm{k}_j)^2}\right]\nonumber\\
&=&
S_2(A)-\frac{1}{d}\sum_j r_j\frac{\bm{k}_j^tC^tC\bm{k}_j}
{1+\bm{r}_B\cdot\bm{k}_j}\,,\label{SAB2f}\end{eqnarray}
where $S_2(A)=S_2(\rho_A)=1-(1+|\bm{r}_A|^2)/d_A$
and  $\bm{k}_j^tC^tC\bm{k}_j=|C\bm{k}_j|^2$, with
\begin{equation}
C=J-\bm{r}_A\bm{r}_B^t=\langle\bm{\sigma}_A\otimes\bm{\sigma_B^t}
\rangle-\langle\bm{\sigma}_A\rangle\langle\bm{\sigma}_B^t\rangle\,,
 \label{CC}\end{equation}
the {\it correlation matrix}, of elements $C_{ij}=\langle
\sigma_{Ai}\otimes\sigma_{Bj}\rangle
-\langle\sigma_{Ai}\rangle\langle\sigma_{Bj}\rangle$
($C=0$ iff $\rho=\rho_A\otimes\rho_B$). The second term in (\ref{SAB2f}) is
just the quadratic information gain (i.e., purity increase) (\ref{I21}):
\begin{equation}
I_2(A|B_{\{\Pi_j\}})
=\frac{1}{d}\sum_j r_j\frac{\bm{k}_j^tC^tC
\bm{k}_j}{1+\bm{r}_B\cdot\bm{k}_j}\,.\label{CTC}
\end{equation}
It is then determined by $\bm{r}_B$ and the $(d_B^2-1)\times (d_B^2-1)$
positive semidefinite matrix $C^tC$. We finally note that we may also express
Eqs.\ (\ref{rhod}) and (\ref{roa2}) in terms of the correlation matrix $C$
(rather than $J$) as
\begin{eqnarray}
\rho&=&\rho_A\otimes\rho_B+
\bm{\sigma}_A^t\,C\otimes\bm{\sigma}_B/d\,,\nonumber\\
\rho_{A/\Pi_j}&=&\rho_A+
\frac{\bm{\sigma}_A^t C\bm{k}_j}{d_A(1+\bm{r}_B\cdot\bm{k}_j)}\,,
\end{eqnarray}
with $||\rho-\rho_A\otimes\rho_B||^2={\rm Tr}\,[C^t C]/d=||C||^2/d$.

\subsection{The qudit-qubit case}
We now show that when $B$ is a single qubit, an analytic expression for the
minimum $S_2$ conditional entropy (i.e., for the maximum conditional purity of
$A$) amongst projective local measurements on $B$ can be obtained for any
dimension $d_A$ of $A$ (${\mathbb{C}}^{d_A}\otimes {\mathbb{C}}^{2}$ system)
and any initial state $\rho$. Here we can take $\bm{\sigma}_B$ as the Pauli
operators, and $\bm{k}_j$ become unit vectors. For a projective spin
measurement along direction $\bm{k}$ ($|\bm{k}|=1$), we have $j=1,2$, with
$r_j=1$, $\bm{k}_1=-\bm{k}_2=\bm{k}$, and Eq.\ (\ref{CTC}) becomes
\begin{eqnarray}
I_2(A|B_{\bm{k}})&=&
\frac{1}{d_A}\frac{\bm{k}^tC^tC\bm{k}}
{1-(\bm{r}_B\cdot\bm{k})^2}=\frac{1}{d_A}
\frac{\bm{k}^tC^tC\bm{k}}{\bm{k}^tN_B\bm{k}}
\,,\label{MC}
\end{eqnarray}
where $N_B$ is the  $3\times 3$ positive semidefinite matrix
\begin{equation} N_B=I_3-\bm{r}_B\bm{r}_B^t\,.\end{equation}
The last expression in (\ref{MC}) is a ratio of quadratic forms and is then
independent of the length of $\bm{k}$. Its maximum can therefore be obtained
diagonalizing the $3\times 3$ matrix $C^tC$ with the metric $N_B$: Setting
$\bm{k}=N_B^{-1/2}\bm{\tilde{k}}$, with $\bm{\tilde{k}}^t\bm{\tilde{k}}=1$, we
have
\begin{eqnarray}\frac{\bm{k}^t C^tC\bm{k}}{\bm{k}^t N_B\bm{k}}&=&\bm{\tilde{k}}^t
 \tilde{C}^t\tilde{C}\bm{\tilde{k}}\leq \lambda_{\rm max}\,,\end{eqnarray}
where $\tilde{C}=C N_B^{-1/2}$ and $\lambda_{\rm max}$ is the maximum
eigenvalue of $\tilde{C}^t\tilde{C}$, the maximum reached when $\bm{\tilde{k}}$
is the associated normalized eigenvector. The eigenvalue equation
$\tilde{C}^t\tilde{C}\bm{\tilde{k}}=\lambda\bm{\tilde{k}}$ is just the
eigenvalue equation for $C^tC$ with metric $N_B$,
\begin{equation}C^tC\bm{k}=\lambda N_B\bm{k}\,,
\end{equation}
so that $\lambda_{\rm max}$ is the largest root of the equation
\begin{equation}{\rm Det}[C^tC-\lambda N_B]=0\,,\end{equation}
with $\bm{k}$ the associated eigenvector. In other words, $\sqrt{\lambda_{\rm
max}}$ is the maximum {\it singular value} of the matrix $C$ with metric $N_B$.
The ensuing minimum conditional entropy and maximum information gain
(uncertainty reduction) for projective measurements are then
\begin{eqnarray}S_2(A|B)&=&\mathop{\rm Min}_{\bm{k}} S_2(A|B_{\bm{k}})
 =S_2(A)-\lambda_{\rm max}/d_A\,,\label{rs}\\
I_2(A|B)&=&\mathop{\rm Max}_{\bm{k}} I_2(A|B_{\bm{k}})
 =\lambda_{\rm max}/d_A\,.\label{rss}\end{eqnarray}
If $\bm{r}_B=\bm{0}$, $N_B=I_3$ and $\lambda_{\rm max}$ is just the maximum
eigenvalue of $C^tC$. On the other hand, if  $|\bm{r}_B|=1$, $\rho$ is a
product state and $\bm{k}^t C^tC\bm{k}=0$ vanishes $\forall$ $\bm{k}$.

For instance, the classically correlated state (\ref{rk}) corresponds, choosing
the $z$ axis in $B$ such that $\tilde{\Pi}^B_{\pm \bm{k}}=\frac{1}{2}(I_B\pm
\sigma_z)$,  to $(\bm{r}_B)_\nu=\delta_{\nu z}r_B$, $J_{\mu\nu}=\delta_{\nu
z}J_{\mu z}$, implying $(C^tC)_{\nu\nu'}=\delta_{\nu\nu'}\delta_{\nu
z}|\bm{J}-r_B\bm{r}_A|^2$, with $\bm{J}$ the vector of components $J_{\mu z}$.
Hence,
\begin{equation}
\frac{\bm{k}^t C^tC\bm{k}} {\bm{k}^t N_B\bm{k}}\leq \lambda_{\rm max}=
 \frac{|\bm{J}-r_B\bm{r}_A|^2}{1-r_B^2}\,,\end{equation}
being verified that the maximum is reached for $\bm{k}$
along $z$, i.e., for a spin measurement along $\bm{r}_B$ (basis of eigenstates
of $\rho_B$). For a general state however, the minimizing direction may differ
from $\bm{r}_B$ and follow the main correlation in $C^tC$.

If $A$ is also a qubit ($d_A=2$), it is convenient to use $S_2(\rho)=2(1-{\rm
Tr}\,\rho^2)$ in previous equations, i.e. $\frac{1}{d_A}\rightarrow 1$ in Eqs.\
(\ref{MC})--(\ref{rss}), such that $S_2(\rho_A)=1$ if $\rho_A$ is maximally
mixed. Such rescaled entropy is still a lower bound to the $a=2$ von Neumann
entropy $S(\rho)=-{\rm Tr}\rho\log_2\rho$ (see \ref{ApA}). In such a case, if
$\rho$ is of rank $2$, it can be purified by adding a third qubit $C$, being
then verified that $S_2(A|B)$ coincides with the {\it squared concurrence}
\cite{WW.97} between $A$ and $C$, since such quantity reduces for pure
two-qubit states to the present rescaled $S_2$ entropy of any of the
subsystems, and coincides with its convex roof extension $E_2(A,C)$ for mixed
two qubit states \cite{Ca.03}.

We remark finally that for a qudit-qubit state, the (minimum) geometric discord
(\ref{D2}) is determined by the largest eigenvalue of a different $3\times 3$
matrix \cite{DVB.10,Mo.12}:
\begin{equation}
D_2(A|B)=\frac{1}{d}(|\bm{r}_B|^2+ ||J||^2-\tilde{\lambda}_{\rm
 max})\,,\label{D22}\end{equation}
where $\tilde{\lambda}_{\rm max}$ is the largest eigenvalue of
$M_2=\bm{r}_B\bm{r}_B^t+J^tJ$. This matrix depends then on $J$ rather than the
correlation $C$, coinciding with $C^tC$ just when $r_B=0$.

 \section{Application}
\subsection{X states}
Let us now consider a two-qubit system. Through its singular value
decomposition, the now $3\times 3$ matrix $J$ can be always brought to the
diagonal form $J_{\mu\nu}=\delta_{\mu\nu} J_\mu$ by appropriately choosing the
local $x,y,z$ axes. If $\bm{r}_A$ and $\bm{r}_B$ are directed along the same
principal axes of $J$, which we shall denote as $z$, we obtain an $X$ state
\cite{AR.10},
\begin{eqnarray}\rho&=&\frac{1}{4}(I+r_A\,\sigma_{z}\otimes I_2+
r_B\,I_2\otimes\sigma_{z}+\!\!\!
\sum_{\mu=x,y,z}\!\!\! J_\mu \sigma_{\mu}\otimes\sigma_{\mu})\nonumber\\&&\label{X}\\
&=&\left(\begin{array}{cccc}p_+&0&0&\alpha_-\\0&q_+&\alpha_+&0\\0&\alpha_+&
q_-&0\\\alpha_-&0&0&p_-\end{array}\right)\,,\;
\begin{array}{c}
p_{\pm}=\frac{1\pm(r_A+r_B)+J_z}{4}\\
q_{\pm}=\frac{1\pm(r_A-r_B)-J_z}{4}\\
\alpha_{\pm}=\frac{J_x\pm J_y}{4}\end{array}\,,\label{m}
\end{eqnarray}
where Eq.\ (\ref{m}) is its standard basis representation. This state commutes
with the $z$ parity $P_z=\sigma_{z}\otimes \sigma_{z}$. Accordingly, reduced
states of arbitrary spin pairs in the thermal state or in any non-degenerate
eigenstate of any spin $1/2$ array with $XY$ or $XYZ$ Heisenberg couplings of
arbitrary range in a field along $z$, are of the present form \cite{CRC.10}, as
the corresponding Hamiltonian (see Eq.\ (\ref{Hxy})) commutes with the total
$z$ parity.

The  ensuing matrices $C$ and $N_B$ are simultaneously diagonal,
\[C=\left(\begin{array}{ccc}J_x&0&0\\0&J_y&0\\0&0&
J_z-r_A r_B\end{array}\right),\;N_B= \left(\begin{array}{ccc}
 1&0&0\\0&1&0\\0&0&r^2_B\end{array}\right)\,. \]
Hence, the minimum conditional entropy $S_2(A|B)$ among projective measurements
will be obtained for a measurement along one of the principal axes $x,y,z$. We
then obtain
\begin{eqnarray}
S_2(A|B)&=&1-|\bm{r}_A|^2-I_2(A|B)\,,\\
I_2(A|B)&=&\mathop{\rm Max}_{\bm{k}}\frac{\bm{k}^tC^tC\bm{k}}{\bm{k}^t
N_B\bm{k}}\nonumber\\&=& {\rm Max}[J_x^2,J_y^2,\frac{(J_z-r_A r_B)^2}{1-r_B^2}]\,,
 \label{I222}\end{eqnarray}
for $S_2(\rho)=2(1-{\rm Tr}\rho^2)$, implying a $z\rightarrow x$ or
$z\rightarrow y$ transition in the direction of the minimizing measurement as
$J_x^2$ or $J_y^2$ increase across  $\lambda_z=(J_z-r_Ar_B)^2/(1-r_B^2)$.

Such direction is then determined essentially by the main correlation in
$C^tC$. This provides a conceptual basis for the results of \cite{LL.11}
related with the minimizing measurement of the {\it quantum discord} for $X$
states, which also follow the main correlation. This direction can then differ
significantly from that minimizing the geometric discord
(\ref{D2})--(\ref{D22}). For the state (\ref{X}) we obtain \cite{DVB.10,RCC.10}
(Eq.\ (\ref{D22}))
\begin{equation}
D_2(A|B)=\frac{1}{2}\{r_B^2+||J||^2-{\rm Max}[J_x^2,J_y^2,J_z^2+r_B^2]\}\,,
\end{equation}
entailing a $z\rightarrow x$ or $z\rightarrow y$ transition only as  $J_x^2$ or
$J_y^2$ increase across $J_z^2+r_B^2$. Coincidence between both minimizing
measurements can then be ensured just for $r_B=0$, i.e., $\rho_B$ maximally
mixed, where the minimizing $\bm{k}$ is along the axis with the largest
$|J_\mu|$ for both $S_2(A|B)$ and $D_2(A|B)$.

For a general entropy $S_f$, the conditional entropy is (Eq.\ (\ref{roa2})),
\begin{equation}
S_f(A|B_{\bm{k}})=\sum_{\mu,\nu=\pm 1}
{\frac{1+\nu\bm{r}_B\cdot\bm{k}}{2}}
\,f[\frac{1}{2}(1+\mu|\bm{r}_A+\frac{\nu C\bm{k}}{1+\nu\bm{r}_B\cdot\bm{k}}|)]\,.
\label{Sff}
\end{equation}
It is verified that for an $X$ state, measurements along any of the principal
axes of $J$ (i.e., $x$, $y$, $z$) are always {\it stationary} ($\delta
S_f(A|B_{\bm{k}})=0$ up to first order in $\delta \bm{k}$), i.e., candidates
for minimizing (\ref{Sff}), although other directions cannot be discarded
(typically in the transitional region between the $z$ and $x$ or $y$ regimes).
On the other hand, for {\it two qubit states with maximally mixed marginals},
which can be written as $X$ states with $r_A=r_B=0$, it is seen from
(\ref{I222}) and (\ref{Sff}) that the minimizing measurement is {\it along the
axis with the largest} $|J_\mu|$, i.e., $\bm{k}$ along the largest eigenvalue
of $J^tJ=C^tC$, {\it for any entropy $S_f$} (universal minimum).

We finally mention that the geometric discord $D_2(A|B)$ was shown in
\cite{GA.112} to be an upper bound to the square of the negativity ${\cal
N}(\rho)$, a computable entanglement monotone \cite{Neg}, given for two qubits
by ${\cal N}(\rho)={\rm Tr}|\rho^{T_B}|-1$, with $\rho^{t_B}$ the partial
transpose, both coinciding  for $\rho$ pure. For $X$ states we obtain here a
similar relation between $I_2(A|B)$ and the squared concurrence ${\cal
C}^2(\rho)$,  with both also coinciding when $\rho$ is pure: For the state
(\ref{m}), the concurrence \cite{WW.97} is ${\cal C}(\rho)=2{\rm
Max}[|\alpha_+|-\sqrt{p_+p_-},|\alpha_-|-\sqrt{q_+q_-},0]$, implying ${\cal
C}(\rho)\leq 2{\rm Max}[|\alpha_+|,|\alpha_-|]$ and hence, since
$|\alpha_{\pm}|\leq {\rm Max}[|J_x|,|J_y|]/2$,
\begin{equation}
{\cal C}^2(\rho)\leq {\rm Max}[J_x^2,J_y^2]\leq I_2(A|B)\,.\label{ineq}
\end{equation}

\subsection{Mixture of a pure state with the maximally mixed state}
As a specific example of (\ref{X}), we consider the mixture (\ref{rw}) in the
two qubit case. By suitable choosing the local axes, we may always write it as
\begin{equation}
\rho=w|\Psi\rangle\langle\Psi|+(1-w)I_4/4,\;\;
|\Psi\rangle=\sqrt{q}|00\rangle+\sqrt{1-q}|11\rangle\,,\label{psii}\end{equation}
which corresponds to an $X$ state with
\[r_{A}=r_B=w(2q-1),\;J_x=-J_y=2w\sqrt{q(1-q)}
 \,,\;J_z=w\,.\]
It is then verified that $\frac{(J_z-r_A r_B)^2}{1-r_B^2}-J_x^2
=\frac{w^2(1-w)^2(1-2q)^2} {1-w^2(1-2q)^2}\geq 0\,,$ implying that
$S_2(A|B_{\bm{k}})$ is minimized by a measurement along $z$ (basis of
eigenstates of $\rho_B$), in agreement with the universal minimum for this
state. It is also seen that for $w=1$ ($\rho$ pure), $w=0$ ($\rho$ maximally
mixed) or $q=1/2$ ($|\Psi\rangle$ maximally entangled) the previous difference
vanishes, indicating that all directions $\bm{k}$ lead to the same result, in
agreement with previous considerations. In any case we obtain, for
$S_2(\rho)=2(1-{\rm Tr}\,\rho^2)$,
\begin{eqnarray}S_2(A|B)&=&
\frac{(1-w)(1+w-2w^2(1-2q)^2)}{1-w^2(1-2q)^2}\label{Smin}\,,\\
I_2(A|B)&=&
 \frac{w^2(1-w(1-2q)^2)^2}{1-w^2(1-2q)^2}\label{Smax}\,,\end{eqnarray}
with $S_2(A)=1-w^2(1-2q)^2$.  It is verified that Eq.\ (\ref{Smin}) is a
strictly concave decreasing function of $w$ at fixed $q\in[0,1]$, and a
strictly concave function of $q$ if $w\in(0,1)$, reaching its maximum at
$q=1/2$ (Bell state). Notice that $(1-2q)^2=1-{\cal C}^2(|\Psi\rangle)$, with
${\cal C}(|\Psi\rangle)=2\sqrt{q(1-q)}$ the {\it concurrence} \cite{WW.97} of
$|\Psi\rangle$, so that Eq.\ (\ref{Smin}) is, for $w\in(0,1)$, an increasing
function of ${\cal C}(|\Psi\rangle)$, i.e. of entanglement, as previously
ascertained. The bound  (\ref{ineq}) is also verified (${\cal C}(\rho)={\rm
Max}[w\,{\cal C}(|\Psi\rangle)-(1-w)/2,0]$).

\begin{figure}
\centerline{\scalebox{.5}{\includegraphics{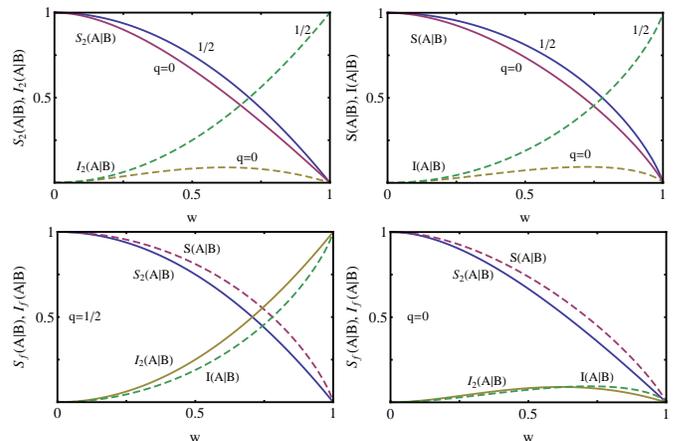}}} \vspace*{-0.5cm}
\caption{(Color online) Top: Results for the quadratic (left) and von Neumann
(right) minimum conditional entropy $S_f(A|B)$ (solid lines) and maximum
information gain (or uncertainty reduction) $I_f(A|B)$
(dashed lines), after a measurement
at $B$ in the mixture (\ref{psii}) for the maximally
entangled ($q=1/2$) and separable ($q=0$) cases. All $S_f(A|B)$ are concave
decreasing functions of $w$, vanishing at the pure limit $w=1$. Bottom:
Comparison between quadratic (solid lines) and von Neumann  (dashed
lines) results for $q=1/2$ (left) and $q=0$ (right). It is verified that
$S_2(A|B)\leq S(A|B)$ $\forall$ $w,q$.}
 \label{f1}
\end{figure}

Eq.\ (\ref{Smax}) is also a strictly concave function of $q$ if
$w\in(0,1]$, maximum at $q=1/2$, i.e., an {\it increasing function} of the
concurrence ${\cal C}(|\Psi\rangle)$. In contrast, Eq.\ (\ref{Smax}) is not
necessarily an increasing function of $w$. Its behavior with $w$ can be
non-monotonous if $|\Psi\rangle$ is separable or almost separable ($q$ small or
close to $1$), as shown in Fig.\ \ref{f1}, where results for the von Neumann
based ($S(\rho)=-{\rm Tr}\rho\log_2\rho$) conditional entropy and information
gain  are also depicted.  Such behavior is universal, i.e., present for any
$S_f$: When $|\Psi\rangle$ is separable, noise induces a non-zero value of
$I_f(A|B)$, since $\rho$ ceases to be a product state
for $w\in(0,1)$. As seen in Fig.\ \ref{f1}, the qualitative behavior of the
minimum linear  and von Neumann conditional entropies is entirely similar, and
the same holds for the ensuing maximum $I_f(A|B)$. Nonetheless, while
$S_2(A|B)\leq S(A|B)$, there is in general no fixed order relation between
$I_2(A|B)$ and $I(A|B)$.

\subsection{Mixture of aligned states}
\begin{figure}
\centerline{\scalebox{.5}{\includegraphics{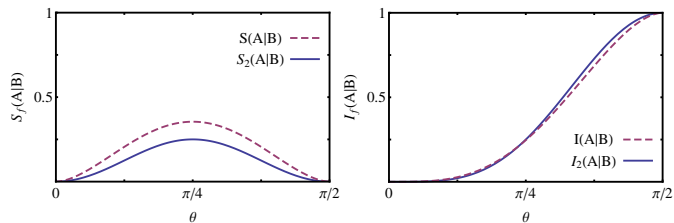}}} \vspace*{-.5cm}
\caption{(Color online) Results for the quadratic (solid lines) and von Neumann
(dashed lines) minimum conditional entropy (left) and maximum information gain
or uncertainty reduction (right) in the mixture of aligned states (\ref{th}).
Both entropies are minimized by a spin measurement along $x$ $\forall$
$\theta\in(0,\pi/2]$.}
 \label{f2}
\end{figure}

We now consider the two-qubit mixed state
\begin{equation}\rho={\textstyle\frac{1}{2}}
(|\theta\theta\rangle\langle\theta\theta|+
 |-\theta\theta\rangle \langle-\theta-\theta|)\,,\label{th}\end{equation}
where $|\theta\rangle=\exp[-i\theta\sigma_y/2]|0\rangle=
\cos\frac{\theta}{2}|0\rangle+\sin\frac{\theta}{2}|1\rangle$ is the state with
the spin forming an angle $\theta$ with the $z$ axis. This separable state
represents, roughly, the reduced state of a spin $1/2$ pair in the exact
definite parity ground state of a ferromagnetic $XY$ chain for fields $|B|<B_c$
if $\cos\theta=B/B_c$ \cite{CRC.10}. Moreover, for not too small chains it is
the {\it exact} state of the pair in the immediate vicinity of the factorizing
field \cite{CRC.10,RCM.08,GAI.08}. Eq.\ (\ref{th}) is an $X$ state with
\[r_A=r_B=\cos\theta,\;\;J_z=\cos^2\theta,\;\;
 J_x=\sin^2\theta,\;\;J_y=0\,.\]
Hence, there is no correlation along $z$ ($J_z=r_A r_B$, implying $C_z=0$) but
there is a finite correlation along $x$ ($C_x=J_x^2$). We then obtain the
remarkable result that $S_2(A|B_{\bm{k}})$ {\it is minimized for $\bm{k}$
along} $x$ $\forall$ $\theta\in(0,\pi/2]$, leading to
\begin{eqnarray}
S_2(A|B)&=&1-\cos^2\theta-\sin^4\theta
={\textstyle\frac{1}{4}}\sin^2\,2\theta\,\label{th1},\nonumber\\
I_2(A|B)&=&\sin^4\theta\,.
 \label{th2}\end{eqnarray}
The minimum $S_2$ conditional entropy is then symmetric around $\theta=\pi/4$,
vanishing for $\theta=0$ (product state) and $\pi/2$ (classically correlated
state of the form (\ref{rk}) with $\rho_{A/\bm{k}}$ pure), whereas the maximum
$I_2(A|B)$ increases with $\theta$ (Fig.\ \ref{f2}), reaching its
absolute maximum at $\theta=\pi/2$.  Hence, spin measurements along $z$ are not
minimum for {\it any} $\theta>0$ (although the difference with (\ref{th1}) is
$O(\theta^4)$ for $\theta\rightarrow 0$).

In the von Neumann case, the behavior of $S(A|B)$ and $I(A|B)$ is again
completely similar to that of $S_2(A|B)$ and $I_2(A|B)$, as seen in Fig.\
\ref{f2}. Moreover, the minimizing measurement is also for $\bm{k}$ along $x$
$\forall \theta\in(0,\pi/2]$ \cite{CRC.10,RCC.10}, i.e., {\it  the same as that
of the $S_2$ entropy} $\forall$ $\theta$. The $S_2$ results allow then to
easily understand the minimizing measurement of the quantum discord for this
state \cite{CRC.10}. In contrast, the geometric discord is minimized for
$\bm{k}$ along $x$ {\it only if} $\theta>\theta_c$, with
$\cos^2\theta_c=\frac{1}{3}$, preferring $\bm{k}$ along $z$ if
$\theta<\theta_c$ \cite{RCC.10}.

\subsection{Spin $1/2$ pairs in $XY$ chains at strong transverse fields}
Let us finally consider a spin $1/2$ array with $XY$ couplings in a strong
transverse field, described by a Hamiltonian
\begin{equation}H=-B\sum_i \sigma_{iz}-\sum_{i<j}(J^x_{ij}
\sigma_{ix}\sigma_{jx}+J^y_{ij}\sigma_{iy}\sigma_{jy})\,.
\label{Hxy}\end{equation}
For sufficiently strong fields $B\gg |J_{ij}^\mu|$ $\forall$ $\mu,i,j$, the
system is weakly coupled and  the ground state is of the form
\begin{equation}|\Psi\rangle\approx |0\rangle+\sum_{i<j}\alpha_{ij}|ij\rangle\,,
\end{equation}
at lowest non trivial order, where $|0\rangle=|0\ldots 0\rangle$ denotes the
state with all spins aligned along the field ($+z$),
$|ij\rangle=\sigma_{i-}\sigma_{j-}|0\rangle$ and $\alpha_{ij}\approx
(J^x_{ij}-J^y_{ij})/(2B)$. The reduced state of a pair $i,j$ is therefore an
$X$ state with, at lowest non-zero order (we set $\alpha_{ji}=\alpha_{ij}$),
\begin{eqnarray}\alpha_-&=&\alpha_{ij},\;\;p_-=|\alpha_{ij}|^2\,,\nonumber\\
\alpha_+&=&\sum_{k\neq
i,j}\alpha_{ik}\bar{\alpha}_{kj}\,,\;\;q_{\pm}=\sum_{k\neq i,j}
 |\alpha_{{^i_j}k}|^2\, .\end{eqnarray}
By suitably choosing the local states at sites $i,j$ we may set $\alpha_{\pm}$
real and positive. Hence, up to $O(|\alpha|^2)$ we obtain
$r_{A,B}=1-2(|\alpha_{ij}|^2+q_{\mp})$ (along $z$) and
\begin{equation}J_{^x_y}=2(\sum_{k\neq i,j}\alpha_{ik}
 \bar{\alpha}_{kj}\pm\alpha_{ij}),\;\;J_z-r_{A}r_B\approx 4|\alpha_{ij}|^2\,.
 \end{equation}
Hence,  for $\alpha_{ij}\neq 0$ (interacting pair), $C_{xx}$ is
$O(\alpha_{ij})$ whereas $C_{zz}$ is $O(\alpha_{ij}^2)$, entailing at lowest
order {\it a minimizing measurement along $x$} instead of $z$, as the
correlation along $z$ is of higher order. {\it The same behavior} occurs with
the minimizing measurement of the von Neumann conditional entropy and hence the
quantum discord in this regime ($\bm{k}$ along $x$  at strong fields
\cite{CRC.10,CCR.13}). In contrast, that minimizing the geometric discord or
the information deficit \cite{RCC.10,SKB.11} follows the main component of the
state, and is therefore along the field direction $z$ for strong fields
\cite{CCR.13}.

\section{Conclusions}
We have analyzed the main features of the conditional entropy associated to
general concave entropic forms in bipartite quantum systems, determined by a
measurement in one of the constituents. Its minimum among all local
measurements determines the maximum average uncertainty reduction (generalized
information gain) about $A$ that can be achieved by a measurement on $B$, and
has the direct meaning of representing the associated entanglement of formation
between $A$ and a purifying third system $C$. For some important classes of
states as those of sections 2.3 and 2.4, the minimizing measurement is {\it the
same} for all $S_f$ and can be analytically and identified, allowing a direct
general evaluation of $E_f(A,C)$. This universality indicates that for such
states there is clearly an unambiguous optimum local measurement leading to the
lowest conditional mixedness at the unmeasured part, irrespective of the
measure used for quantifying such mixedness.

For the general case, a main practical result of our manuscript is the analytic
determination of this minimum for the linear entropy $S_2$ in a general
qudit+qubit state with projective measurements. It can be expressed in terms of
the largest eigenvalue of a simple $3\times 3$ matrix, which represents the
largest singular value of the correlation matrix $C$ with a metric $N_B$
determined by the measured part. This enables to easily identify the minimizing
measurement, determined by the associated eigenvector,  and understand its
behavior. Conditional $S_2$ results have also a direct interpretation in terms
of purity and average distances, and possess the importance of determining the
universal behavior of {\it all} conditional entropies and the ensuing
minimizing measurement in the vicinity of maximum mixedness.

In the specific examples considered, the minimizing measurements of the $S_2$ and von
Neumann conditional entropies (and hence the quantum discord) were in fact
coincident. The present results explain then the quite distinct response of
this minimizing measurement to the onset of correlations (it follows the main
correlation even if arbitrarily weak), in comparison with those minimizing the
geometric discord or the one way information deficit, which follow instead the
main component of the state \cite{CCR.13}. Hence, the present formalism not
only allows to identify universal features and optimize post-measurement
purities, but can also help to evaluate or estimate the quantum discord in more
complex situations, as the minimizing measurements for the linear and von
Neumann conditional entropies become coincident in some states and regimes, and
can be expected to be close in typical situations.

The authors acknowledge support of CIC (RR) and CONICET (NG) of Argentina.

\appendix
\section{\label{ApA} Trace form generalized entropies}

Given a quantum state $\rho$ with spectral decomposition $\rho=\sum_{j}
p_j|j\rangle\langle j|$, $j=1,\ldots,d$ ($p_j\geq 0$, $\sum_{j} p_j=1$), the
``entropic'' forms (see for instance \cite{Wh.78,CR.02})
\begin{equation}
S_f(\rho)={\rm Tr}\,f(\rho)=\sum_j f(p_j)\,,\label{SfAA}
\end{equation}
comply, for any strictly concave real function $f:[0,1]\rightarrow \Re$
satisfying $f(0)=f(1)=0$, with all conventional entropy properties except
additivity: i) $S_f(\rho)\geq 0$, with $S_f(\rho)=0$ iff $\rho$ is pure
($\rho^2=\rho$), ii) $S_f(\rho)$ is maximum at the maximally mixed state
$I_d/d$, with $S_f(I_d/d)=df(1/d)$ an increasing function of $d$, iii)
$S_f(U\rho U^\dagger)=S_f(\rho)$ $\forall$ unitary $U$ and iv) $S_f(\rho)$ is
concave (Eq.\ \ref{conc}) (if $\rho=\sum_\alpha q_\alpha \rho_\alpha$,
$f(p_j)=f(\sum_{\alpha,j'} q_\alpha |\langle j|j'_\alpha\rangle|^2
p_{j'}^\alpha) \geq \sum_{j,'\alpha}q_\alpha |\langle j|j'_\alpha\rangle|^2
f(p_{j'}^\alpha)$, which leads to (\ref{conc}) after summing over $j$).

Concavity implies ii) and, moreover, the majorization \cite{Wh.78,Bh.97}
property \cite{RC.03,RCC.10}
\begin{equation}
\rho\prec\rho'\;\;\Rightarrow S_f(\rho)\geq S_f(\rho')\,,
 \label{maj}\end{equation}
where $\rho\prec\rho'$ ($\rho$ more mixed than $\rho'$) means $\sum_{j=1}^i
p_j\leq\sum_{j=1}^i p'_j$ for $i=1,\ldots,d-1$, with $p_j$, $p'_j$ denoting
here the eigenvalues of $\rho$ and $\rho'$ sorted in {\it decreasing} order
(and completed with $0$'s if dimensions differ). Eq.\ (\ref{maj}) provides the
conceptual basis for considering any such $S_f$ a generalized uncertainty
measure or entropic form. Furthermore, while the converse of (\ref{maj}) does
not necessarily hold if valid for some particular $S_f$ (majorization is
stronger than a single entropic inequality), it {\it does hold} if valid
$\forall$ $S_f$ of the form (\ref{SfAA}): $S_f(\rho)\geq S_f(\rho')$ $\forall$
$S_f$ $\Rightarrow$ $\rho\prec\rho'$ \cite{RC.03}.

The Tsallis entropy \cite{Ts.09} $S_q(\rho)=(1-{\rm Tr}\,\rho^q)/(q-1)$, $q>0$,
corresponds to $f(\rho)=(\rho-\rho^q)/(q-1)$ in  (\ref{SfAA}). It reduces to
the quadratic entropy (\ref{S2}) for $q=2$ and to the von Neumann entropy (with
$a=e$) for $q\rightarrow 1$. We may also set $S_q(\rho)=(1-{\rm
Tr}\,\rho^q)/(1-2^{1-q})$, such that $S_q(\rho)=1$ for a maximally mixed single
qubit state $\rho=I_2/2$, in which case $S_2(\rho)=2(1-{\rm Tr}\rho^2)$ and
$S_q(\rho)\rightarrow -{\rm Tr}\rho\log_2 \rho$ for $q\rightarrow 1$. For this
scaling it is still verified that $S_2(\rho)\leq S(\rho)$ for any single
qubit state, coinciding just for $\rho$ pure or maximally mixed (for any single
qubit state, $S_q(\rho)\leq S(\rho)$ for $1<q<q_1\approx 4.718$ with this scaling).

For two {\it classical} random variables $A,B$ described by a joint probability
distribution $p_{ij}=p(A=i,B=j)$, we may define a generalized conditional entropy
$S_f(A|B)$ as
\begin{equation}
S_f(A|B)=\sum_j p_j S_f(A|B=j)=\sum_{i,j}p_jf(p_{ij}/p_j)\,,
 \label{SABCL}\end{equation}
where $p_j=p(B=j)=\sum_i p_{ij}$. This quantity measures the average
uncertainty about $A$ if $B$ is known. Due to concavity, it satisfies
$S_f(A|B)\leq S_f(A)=\sum_i f(q_{i})$ (with $q_i=p(A=i)=\sum_j p_{ij}$)
$\forall$ $S_f$. The difference
\[I_f(A|B)=S_f(A)-S_f(A|B)\,,\]
is then non-negative, vanishing only if $p_{ij}/p_j=p_i$ $\forall$ $i,j$ with
$p_j>0$, i.e., only if $A$ and $B$ are independent. It represents the
uncertainty reduction (or generalized ``information gain'') about $A$ generated
by the knowledge of $B$.

In the Shannon case $f(p)=-p\log_a p$,  (\ref{SABCL}) becomes
$S(A|B)=S(A,B)-S(B)$, where $S(A,B)=-\sum_{i,j}p_{ij}\log_a p_{ij}$,
$S(B)=-\sum_j p_j\log_a p_j$, but such relation no longer holds for a general
$S_f$. Hence, while in the (classical) Shannon case
$I(A|B)=S(A)+S(B)-S(A,B)=I(B|A)$ is the mutual information, for a general
$S_f$, $I_f(A|B)$ will differ in general from $I_f(B|A)$. Generalizations of
the Shannon conditional entropy based on the Renyi entropy  were recently
discussed  in \cite{Te.12}  (and quantum versions in \cite{ML.13,AR.13}),
whereas special extensions for the Tsallis case were considered in
\cite{AR.01}.

\section{\label{ApB} Relation with the entanglement of formation}
Let us sketch the proof of the identity (\ref{idf})
\cite{KW.04,CC.11,MD.11,FC.11}. Starting from the $(AC,B)$ Schmidt
decomposition of the pure global state,
\begin{equation}|\Psi_{ACB}\rangle=\sum_{k=1}^n \sqrt{q_k}\,
|\tilde{k}_{AC}\rangle|\tilde{k}_B\rangle\label{SAC}\,,
\end{equation}
the state of $AC$ after a measurement in $B$ based on the operators (\ref{M1})
with outcome $j$ is the pure state (Eq.\ (\ref{ja}))
\begin{equation} |j_{AC}\rangle=(r_j/p_j)^{1/2}\sum_{k}\sqrt{q_k}
\langle
 j_B|\tilde{k}_B\rangle|\tilde{k}_{AC}\rangle\,.\label{jac}\end{equation}
Hence, $\rho_{A/\Pi_j}$ is the reduced state $\rho_A^j$ of $A$ in
$|j_{AC}\rangle$ and $S_f(A|B_{\{\Pi_j\}})=\sum_j p_j S_f(\rho^j_{A})$
coincides then with the average entanglement of the decomposition
$\rho_{AC}=\sum_j p_j \rho_{AC/\Pi_j}$, where
$\rho_{AC/\Pi_j}=|j_{AC}\rangle\langle j_{AC}|$. Conversely, Eq.\ (\ref{SAC})
implies that the  states $|j_{AC}\rangle$ in any  decomposition
$\rho_{AC}=\sum_j p_j |j_{AC}\rangle\langle j_{AC}|$ (with $p_j>0$) should
satisfy
\begin{equation}\sqrt{p_j}|j_{AC}\rangle=\sum_k U_{jk}
\sqrt{q_k}|\tilde{k}_{AC}\rangle\,,\end{equation} where $U$ is an $m\times n$
matrix with orthonormal columns ($\sum_j U^*_{jk}U_{jk'}=\delta_{kk'}$) and
$m\geq n$. Comparison with Eq.\ (\ref{jac}) indicates that we may identify such
decomposition with that for a local measurement at $B$ with the operators
(\ref{M1}),  provided
\begin{equation}
 \sqrt{r_j}|j_B\rangle=\sum_k U^*_{jk}|\tilde{k}_B\rangle\,,\end{equation}
such that $U_{jk}=\sqrt{r_j}\langle j_B|\tilde{k}_B\rangle$. The ensuing
operators $\Pi_j^B=r_j|j_B\rangle\langle j_B|$ form a valid POVM since $\sum_j
\Pi_j^B=\sum_{j,k,k'} U_{jk}^*U_{jk'}| \tilde{k}_B\rangle\langle
\tilde{k}'_B|=\sum_k |\tilde{k}_B\rangle\langle \tilde{k}_B|=I_B$ (assuming
$n=d_B$).

\end{document}